\documentclass[prl,aps,amssymb,twocolumn,showpacs,floatfix]{revtex4}
\usepackage[dvips]{graphicx}
\def\be{\begin{equation}}
\def\ee{\end{equation}}
\def\bea{\begin{eqnarray}}
\def\eea{\end{eqnarray}}

\begin{document}
 
\title{Dephasing and weak localization in disordered Luttinger liquid}
\author{I.V.~Gornyi$^{1,*}$}
\author{A.D.~Mirlin$^{1,2,\dagger}$}
\author{D.G.~Polyakov$^{1,*}$}
\affiliation{$^{1}$Institut f\"ur Nanotechnologie,
Forschungszentrum Karlsruhe, 76021 Karlsruhe, Germany \\
$^{2}$Institut f\"ur Theorie der kondensierten Materie, Universit\"at
Karlsruhe, 76128 Karlsruhe, Germany}
 
\date{\today}

\begin{abstract} We study the transport properties of interacting electrons in
a disordered quantum wire within the framework of the Luttinger liquid
model. The conductivity at finite temperature is nonzero only because of
inelastic electron-electron scattering. We demonstrate that the notion of weak
localization is applicable to the strongly correlated one-dimensional electron
system. We calculate the relevant dephasing rate, which for spinless electrons
is governed by the interplay of electron-electron interaction and disorder,
thus vanishing in the clean limit. \end{abstract}
 
 
\maketitle

Mesoscopics of strongly correlated electron systems has emerged as an area of
great interest to both experimental and theoretical communities working in the
field of nanoscale physics. Recently, progress in manufacturing of nanodevices
has paved the way for systematic transport measurements on narrow quantum
wires with a few or single conducting channels. Most prominent examples of
these are semiconductor cleaved-edge quantum wires \cite{auslaender02}, carbon
nanotubes \cite{man04}, and quantum Hall edges running in opposite directions
and interconnected by means of tunneling \cite{kang00,grayson05}. On the
theoretical side, the challenge is to expand the ideas that have been
developed for mesoscopic disordered systems on one side and for strongly
correlated clean systems on the other.

Much attention has been focused on the interplay between the interaction
effects and disorder-induced localization in diffusive systems of low
dimensionality $D$ \cite{altshuler85}. A key concept in the localization
theory of a disordered Fermi liquid is that of the dephasing rate
$\tau_\phi^{-1}$ due to electron-electron (e-e) inelastic scattering. It has
been established that a weak-localization (WL) correction to the Drude
conductivity of a diffusive system behaves as $\tau_\phi^{(2-D)/2}$
($\ln\tau_\phi$ for $D=2$) and thus diverges with lowering $T$ for $D\leq 2$,
leading to strong Anderson localization.
 
This paper is concerned with transport in one dimension (1D), where e-e
correlations drive a clean system into the non-Fermi liquid state known as
Luttinger liquid (LL) \cite{giamarchi04}. One more peculiarity of the
single-channel 1D system is that the ballistic motion on short scales crosses
over in the absence of interaction directly to the localization regime, with
no diffusive dynamics on intermediate scales. The main question we address is
how the conductivity $\sigma (T)$ behaves in a disordered LL. It appears that
a key piece of transport theory as regards the WL and the interaction-induced
dephasing in a strongly correlated 1D system is missing. Most authors to date
(e.g., \cite{giamarchi88,furusaki93}) have suggested that the dephasing length
that controls localization effects in a disordered LL is $L_T=u/T$ (throughout
the paper $\hbar=1$), where $u$ is the plasmon velocity. According to this
approach, the interference effects get strong with lowering $T$ at
$L_T\sim\xi$, where $\xi$ is the localization length. An alternative approach
\cite{apel82a,lehur02} is predicated on the assumption that the dephasing rate
is determined by the single-particle properties of a clean LL. On top of that,
one might think that since in the case of linear dispersion the interacting
electron system can be equivalently represented in terms of {\it
noninteracting} bosons, the interaction should not induce any dephasing at
all. The conductivity would then be exactly zero at any $T$. As we argue
below, none of the approaches captures the essential physics of dephasing in
the conductivity of a disordered 1D system.

We begin by considering the Drude conductivity under the condition that
$\tau_\phi$ is much shorter than the transport time of elastic
scattering off disorder $\tau$ and the Anderson-localization effects are
completely destroyed. For simplicity, we assume that interaction is weak and
short-ranged. We also assume that $\epsilon_F\tau\gg 1$, where $\epsilon_F$ is
the Fermi energy. To leading order in $\tau_\phi/\tau\ll 1$, the conductivity
is given by the Drude formula $\sigma^{\rm D}=e^2\rho v_F^2\tau$
($\rho=\partial n/\partial\mu\simeq 1/\pi v_F$ is the compressibility, $v_F$
the Fermi velocity) and depends on $T$ through a $T$-dependent renormalization
of the static disorder \cite{mattis74,giamarchi88}:
\begin{equation}
\tau_0/\tau=(\Lambda/T)^{2\alpha'}~,
\label{1} 
\end{equation} 
where $\alpha'=\eta_s^{-1}[1-(1+2\eta_s\alpha)^{-1/2}]\simeq \alpha=V_f/2\pi
v_F>0$ characterizes the strength of repulsive interaction between electrons
(we assume that $\alpha\ll 1$), $\eta_s=1$ or 2 for spinless or spinful
electrons, respectively; $V_f$ is the Fourier transform of a
forward-scattering potential, $\tau_0^{-1}$ the scattering rate at
$\alpha=0$. For $\alpha\ll 1$, the ultraviolet cutoff $\Lambda$ may be put
equal to $\epsilon_F$. The exponent in Eq.~(\ref{1}) is given by the bare
interaction constant (the one in a clean system) since the running coupling
constant \cite{giamarchi88} is not renormalized by disorder for $T\tau\gg
1$. The renormalization (\ref{1}) is similar to the Altshuler-Aronov
corrections \cite{altshuler85} in higher dimensionalities. At
this level, the only peculiarity of LL as compared to higher $D$ is that the
renormalization of $\tau$ is more singular and necessitates going beyond the
Hartree-Fock (HF) approach \cite{polyakov03}.

The renormalization of $\tau$ stops with decreasing $T$ at $T\tau\sim 1$. This
condition gives the zero-$T$ localization length $\xi\propto
\tau_0^{1/(1+2\alpha')}$, but does not correctly predict the onset of
localization (determined by the condition $\tau/\tau_\phi\sim 1$), in contrast
to the argument made in Refs.~\onlinecite{giamarchi88,furusaki93}. This can be
seen, in particular, by noting that the temperature $T\sim \tau^{-1}$ does not
depend on the strength of interaction for small $\alpha$, whereas it is
evident that for noninteracting electrons $\sigma(T)=0$ for any $T$. The error
appears to be based on the renormalization-group equations \cite{giamarchi88},
which treat scalings with length and $u/T$ as interchangeable. While this
approach is justified for the ``elastic renormalization" of $\tau$,
Eq.~(\ref{1}), it does not properly account for the WL and misses all effects
associated with dephasing by construction.

Let us now turn to the calculation of $\tau_\phi^{-1}$. Our approach
is closely related to that for higher dimensionalities \cite{altshuler85} and
it is instructive to first analyze the Golden Rule expression for the
e-e collision rate following from
the Boltzmann kinetic equation:
\begin{eqnarray}
{1\over\tau_{\rm ee} (\epsilon)}=\!\!\int\! \!d\omega\! 
\int \!\!d\epsilon' 
K_\omega
(f^h_{\epsilon-\omega}
f_{\epsilon'}
f^h_{\epsilon'+\omega}+f_{\epsilon-\omega}
f^h_{\epsilon'}
f_{\epsilon'+\omega}),
\label{5}
\end{eqnarray}
where $K_\omega$ is the kernel of the e-e collision integral,
$f_\epsilon$ is the Fermi distribution function, and
$f^h_\epsilon=1-f_\epsilon$. Peculiar to 1D are highly singular contributions
to $K_\omega$ related to scattering of electrons moving in the same
direction. Indeed, consider a perturbative expansion of $K_\omega$ to second
order in $\alpha$ in a clean LL. For simplicity, let $\alpha$ be a
momentum-independent constant. At the Fermi level $(\epsilon=0)$,
$1/2\tau_{\rm ee}=\eta_s(\Sigma^H_{++}+\Sigma^H_{+-})+\Sigma^F$, where
\begin{equation} 
\Sigma^H_{+\pm}\simeq
\pi\alpha^2v_FT\int_{|\omega|\alt T}\!\!\! d\omega\!\int \! dq\,
\delta(\omega-v_Fq)\delta(\omega\mp v_Fq)
\label{6}
\end{equation} 
are the Hartree terms for scattering of two electrons from the same (++) or
different $(+-)$ chiral spectral branches and $\Sigma^F=-\Sigma^H_{++}$ is the
exchange term.  One sees that the contribution of $\Sigma^H_{++}$ is
diverging. For spin-polarized electrons it is, however, canceled by the
exchange interaction. The remaining term $\Sigma^H_{+-}$ is determined by
$\omega,q\to 0$ and is given by $2\Sigma^H_{+-}=\pi \alpha^2T$. Already the
perturbative expansion demonstrates a qualitative difference between two cases
of spinless and spinful electrons.

Below we concentrate on the spinless case. Terms of higher order in $\alpha$
may then be neglected due to the order-by-order cancellation of the singular
Hartree and exchange contributions, so that we obtain
\begin{equation}
\tau_{\rm ee}^{-1}=\pi\alpha^2 T~.
\label{7}
\end{equation}
It is instructive to compare this collision rate in a clean LL with the
damping of the retarded single-particle Green's function in the $(x,t)$
representation, $g^R(x,t)=2i\theta(t){\rm Im}\,\bar{g}(x,t)$, where (for
right-movers) 
\[
\bar{g}={T/2u\over\sinh[\pi T({x\over u}-t+i0)]}
{(\pi
T/\Lambda)^{\alpha_b} \over [\sinh(\pi
T\tau_-)\sinh(\pi T\tau_+)]^{\alpha_b/2}}~,
\]
$\tau_\pm=\pm(t-i/\Lambda)+x/u$, and
$\alpha_b=[(1+2\alpha)^{1/4}-(1+2\alpha)^{-1/4}]^2/2\simeq \alpha^2/2$ for
$\alpha\ll 1$. The temporal decay $\exp (-\pi\alpha_bTt)$ of the residue
\begin{equation}
(x-ut)g^R(x,t)|_{x\to ut}\propto \sinh^{-\alpha_b/2}(2\pi Tt)
\label{8a}
\end{equation}
for $t\to\infty$ agrees with Eq.~(\ref{7}) to order $\alpha^2$: $\pi\alpha_bT
=1/2\tau_{\rm ee}$. The e-e scattering thus manifests
itself in that it cuts off the power-law decay in Eq.~(\ref{8a}),
characteristic of the zero-$T$ limit.

The notion of dephasing associated with the behavior of the single-particle
Green's function (\ref{8a}) makes sense in a clean LL in the ring geometry,
where this kind of dephasing governs the decay rate $(\tau_\phi^{\rm
AB})^{-1}$ of Aharonov-Bohm (AB) oscillations \cite{lehur02,unpub}. However,
as far as $\sigma(T)$ is concerned, the significance of the dephasing rate in
the limit of high $T$ is that it cuts off a WL correction $\sigma^{\rm wl}$ to
the Drude conductivity \cite{ludwig04}. At this point, it is important to note
that the characteristic energy transfer in Eq.~(\ref{5}), $\omega_0\sim
\tau^{-1}$, is much smaller than $\tau_{\rm ee}^{-1}$ in the WL regime. It
suggests that the dephasing rate $1/\tau_\phi^{\rm wl}$ that determines
$\sigma^{\rm wl}$ requires a self-consistent cutoff in Eq.~(\ref{5}) at
$\omega\sim 1/\tau^{\rm wl}_\phi$ (since soft inelastic scattering with
$qv_F,\omega\ll 1/\tau^{\rm wl}_\phi$ does not affect $\sigma^{\rm wl}$
\cite{altshuler82}), and so is parametrically different from the one in
Eq.~(\ref{7}), $\tau_\phi^{\rm wl}\neq\tau_{\rm ee}$.

\begin{figure}[ht]
\centerline{
\includegraphics[width=7cm]{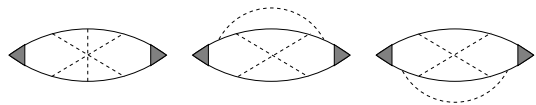}} 
\caption{Diagrams describing the leading WL correction to the conductivity of
Luttinger liquid for $\tau_\phi^{\rm wl}\ll \tau$. The dashed lines represent
backscattering off impurities. The current vertices are dressed by impurity
ladders. The diagrams are understood as fully dressed by e-e interactions. }
\label{f1} 
\end{figure}

To evaluate $\sigma^{\rm wl}$ quantitatively, we use a path-integral
representation: the leading localization correction in the ballistic limit
$\tau_\phi^{\rm wl}/\tau\ll 1$ is related to the interference of electrons
scattered by three impurities. The corresponding diagrams are given by a
``three-impurity Cooperon" (Fig.~\ref{f1}), which describes the propagation of
two electron waves along the path connecting three impurities (``minimal
loop") in time-reversed directions. In the absence of interaction, quantum
interference processes involving a larger number of impurities sum  to
exactly cancel the Drude conductivity \cite{berezinskii73}, which spells
complete localization. For $\tau_\phi^{\rm wl}/\tau\ll 1$, they only yield
subleading corrections through a systematic expansion in powers of
$\tau_\phi^{\rm wl}/\tau$.

\begin{figure}[ht]
\centerline{
\includegraphics[width=7cm]{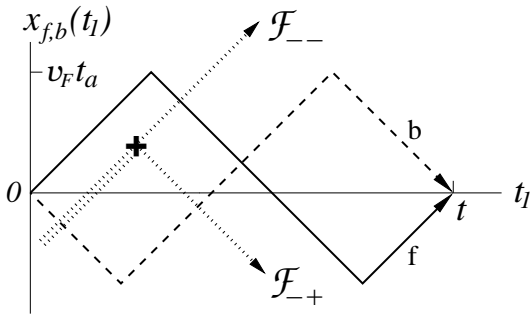}} 
\caption{Illustration of electron dynamics governing the WL and dephasing:
Time-reversed paths $x_f(t_1)$ (solid) and $x_b(t_1)=x_f(t-t_1)$ (dashed) on
which the interaction-induced action $S$ that yields dephasing of the Cooperon
is accumulated. Dotted lines: the propagation of dynamically screened
interaction. The interaction may change the direction of propagation upon
scattering off disorder (as marked by a cross). Each interaction line gives a
contribution to $S$ proportional to $(N_f-N_b)^2$, where $N_{f,b}$ is the
number of its intersections with the forward ($f$) and backward ($b$)
paths. One sees that $N_f\neq N_b$ only due to impurity scattering in the
interaction propagator. Interaction and electron lines lying on top of each
other do not yield dephasing because of the HF cancellation.}
\label{f2} 
\end{figure}

The dephasing-induced action $S(t,t_a)$ acquired by the Cooperon is
accumulated on the classical (saddle-point) path, whose geometry for three
impurities if fixed by two length scales, the total length of the path $v_Ft$
and the distance between two rightmost impurities $v_Ft_a\leq v_Ft/2$
(Fig.~\ref{f2}). The WL correction can then be represented as 
\begin{equation}
\sigma^{\rm wl}=-2\sigma^{\rm D}\!\int_0^\infty \!\!dt \!\int_0^\infty
\!\!dt_a P(t,t_a)\exp \left[-S(t,t_a)\right]~, 
\label{9} 
\end{equation} 
where $P(t,t_a)=(1/8\tau^2)\exp (-t/2\tau)\theta(t-2t_a)$ is the probability
density of return to point $x=0$ after two reflections at points $x=v_Ft_a$
and $x=-v_F(t/2-t_a)$. The contribution $S_{ij}$ to the dephasing action
associated with inelastic interaction between electrons propagating along the
paths $x_i(t)$ and $x_j(t)$ is obtained similarly to higher dimensionalities
\cite{altshuler82,altshuler85}:
\begin{eqnarray} 
S_{ij}=-T\int
{d\omega\over 2\pi}\int {dq\over 2\pi}\int_0^t \!dt_1\!\int_0^t \!dt_2
\,{1\over \omega}{\rm Im}V_{\mu\nu}(\omega,q) \nonumber \\ \times\exp
\{iq\left[\,x_i(t_1)-x_j(t_2)\,\right]-i\omega(t_1-t_2)\}~.  
\label{10}
\end{eqnarray} 

The main steps in the derivation of Eq.~(\ref{10}) are: (i) the random-phase
approximation (RPA), (ii) the independent averaging of each of the RPA bubbles
over disorder, and (iii) treatment of thermal electromagnetic fluctuations
through which electrons interact with each other as a classical field. This
approach is justified if the characteristic energy transfer is much smaller
than $T$, which is the case for $T\tau_\phi^{\rm wl}\gg 1$. Because of the
HF-cancellation of the bare interaction between electrons from the same chiral
branch, the dynamically screened (retarded) interaction $V(\omega,q)$ in
Eq.~(\ref{10}) should be calculated as if the bare coupling is only present
for electrons moving in opposite directions \cite{geology}. As a result,
$V_{\mu\nu}$ acquires the chiral indices $\mu={\rm sgn}\,\dot{x}_i$, $\nu={\rm
sgn}\,\dot{x}_j$. Expanding $V_{\mu\nu}$ to second order in $\alpha$ we have
${\rm Im}V_{\mu\nu}=-\pi\alpha^2v_F\omega {\cal F}_{\mu\nu}$, where ${\cal
F}_{\mu\nu}=4{\rm Re}\,{\cal D}_{-\mu,-\nu}$ and ${\cal D}_{\mu\nu}$ are the
particle-hole propagators for noninteracting electrons. The action $S_{ij}$
can then be written in a simple form:
\[
S_{ij}=\pi\alpha^2v_FT\int_0^t\!\!dt_1\!\int_0^t\!\!dt_2\,
{\cal F}_{\mu\nu}[\,x_i(t_1)-x_j(t_2),t_1-t_2\,]~,
\]
where, to first order in $\tau^{-1}$, ${\cal F}_{\mu\nu}(x,t)$
read
\begin{eqnarray}
{\cal F}_{++}(x,t)&\simeq&\delta(x+v_Ft)\,(1-|t|/2\tau)~,\\
{\cal F}_{+-}(x,t)&\simeq&
\theta(v_F^2t^2-x^2)/4v_F\tau~,
\label{12}
\end{eqnarray}
and ${\cal F}_{--}(x,t)={\cal F}_{++}(-x,t)$, ${\cal F}_{-+}(x,t)={\cal
F}_{+-}(x,t)$. The total action is given by $S=2(S_{\rm ff}-S_{\rm fb})$,
where $f$ and $b$ stand for ``forward" and ``backward" time-reversed paths
(Fig.~\ref{f2}).

Calculating first $S$ for $\tau^{-1}=0$ we have $S_{\rm ff}=S_{\rm
fb}=\pi\alpha^2 Tt/2$. One sees that $S_{\rm ff}$ reproduces the AB dephasing,
Eqs.~(\ref{7}),(\ref{8a}). The subtle point, however, is that for
$\tau^{-1}=0$ the self-energy processes $(S_{\rm ff}+S_{\rm bb})$ are exactly
canceled in $S$ by the vertex corrections $(S_{\rm fb}+S_{\rm bf})$, i.e.,
$S=0$ in a clean LL. Hence, the dephasing in Eq.~(\ref{9}) is only due to the
dressing of the dynamically screened interaction by impurities. To order
$S\sim {\cal O}(\tau^{-1})$ we obtain
\begin{equation}  
S(t,t_a)=2\pi\alpha^2 \,T\,t_a \left(t-2t_a\right)/\tau~.
\label{13}
\end{equation}
The dephasing vanishes for $t_a=0,t/2$ since in these cases the Cooperon is
not distinguishable from the diffuson.

\begin{figure}[ht]
\centerline{
\includegraphics[width=7cm]{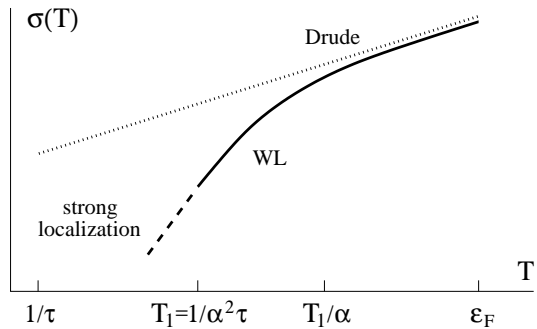}} 
\caption{Schematic behavior of $\sigma (T)$ on the log-log scale. Dotted line:
the $T$-dependent Drude conductivity \cite{giamarchi88,mattis74}. Below
$T_1/\alpha$ the WL correction, Eq.~(\ref{14}), dominates $d\ln\sigma/d\ln
T$. Below $T_1$ the localization becomes strong \cite{unpub}.}
\label{f3} 
\end{figure}

Substituting Eq.~(\ref{13}) in Eq.~(\ref{9}) we find for $\tau_\phi^{\rm
wl}/\tau\ll 1$:
\begin{equation}
\sigma^{\rm wl}=-{1\over 4}\,\sigma^{\rm
D}\left(\tau_\phi^{\rm wl}\over \tau\right)^2\!\ln{\tau\over 
\tau_\phi^{\rm wl}}
\propto {1\over \alpha^2T}\,\ln(\alpha^2T)~,
\label{14}
\end{equation}
where
\begin{equation}
{1\over \tau_\phi^{\rm wl}}=\alpha\left({\pi T\over \tau}\right)^{1/2}~,\quad
T\gg T_1={1\over \alpha^2\tau}~.
\label{15}
\end{equation}
Note that $1/\tau_\phi^{\rm wl}$ vanishes in the clean limit \cite{curvature},
in contrast to the total e-e scattering rate, Eq.~(\ref{7}). It is worth
mentioning that the $T$ dependence of $\sigma(T)=\sigma^{\rm D}(T)+\sigma^{\rm
wl}(T)$ is dominated by the WL term rather than by $\sigma^{\rm D}(T)$ for
$T\ll T_1/\alpha$. The scale $T_1$ marks the temperature below which the
localization effects become strong. These results are illustrated in
Fig.~\ref{f3}.

Before closing the paper, let us briefly mention a few extensions \cite{unpub}
of our results.

(i) {\it ``All-in-one" approach.} In effect, Eqs.~(\ref{14}),(\ref{15}) were
derived in two steps: first the static disorder was renormalized by virtual
processes with energy transfers in the range between $T$ and $\Lambda$ [LL
renormalization, Eq.~(\ref{1})] and then the dephasing rate due to real
processes with transfers smaller than $T$ (similarly to Fermi-liquid
dephasing) was calculated for electrons scattered by the renormalized
disorder.  Alternatively, the virtual and real transitions can be treated on
an equal footing by means of the ``functional bosonization''
\cite{yurkevich}. Including disorder in the bosonic propagators we reproduced
Eqs.~(\ref{14}),(\ref{15}) by this method as well.

(ii) {\it Spin.} In contrast to the spinless case, for $\alpha\ll 1$ the main
contribution to $\tau_\phi^{-1}$ of spinful electrons comes from scattering of
electrons from the same chiral branch on each other. In the clean limit, the
perturbative expansion of $\tau_\phi^{-1}$ in powers of $\alpha$ is diverging
at each order, as in Eq.~(\ref{6}). The most singular terms in
$\tau_\phi^{-1}$ can be summed by means of the RPA and written after the HF
cancellation in the form of $2\Sigma_{++}^H$, Eq.~(\ref{6}), with $v_F$ in one
of the $\delta$-functions being replaced by the plasmon velocity $u$. Due to
the HF cancellation, the latter is taken here as if electrons were spinless,
i.e., from $(u/v_F)^2=1+2\eta_s\alpha$ with $\eta_s=1$. For $\alpha\ll 1$ this
gives
\begin{equation}
{1\over \tau_\phi}=2\pi\alpha^2{v_F\over |u-v_F|}T\simeq 
2\pi\alpha T~,\quad T\gg
T_1^s={1\over \alpha\tau}~.
\label{16}
\end{equation}
This result agrees with the behavior of $\ln |(x-ut)^{1/2}g^R(x,t)|\to
-t/2\tau_\phi$ at $x=ut$ for spinful electrons in a clean LL, similarly to
Eq.~(\ref{8a}). In contrast to spinless electrons, Eq.~(\ref{16}) describes
the dephasing in both the AB and WL setups; in the latter case, we have
$\sigma^{\rm wl}\sim -\sigma^{\rm D}\tau_\phi/\tau$.  Below $T_1^s$ the
localization sets in.

(iii) {\it Low temperature.} In this paper, we have investigated transport at
sufficiently high $T\gg T_1$, when $\tau^{\rm wl}_\phi/\tau\ll 1$. Below $T_1$
the effects of Anderson localization become strong. With lowering $T$ they
lead first to an intermediate regime of ``power-law hopping" \cite{gogolin75},
where $\sigma(T)\sim \sigma^{\rm D}\tau/\tau_\phi$ is a power-law function of
$T$. For still lower $T$, the system enters the ``Anderson-Fock glass" phase,
where $\sigma(T)$ vanishes due to the Anderson localization transition in
many-body Fock space \cite{unpub,aleiner05,altshuler97}.

In conclusion, we have studied the dephasing of WL in a disordered LL. For
spinless electrons, our main result is the WL correction (\ref{14}), governed
by the dephasing rate (\ref{15}). The latter is parametrically different from
the AB dephasing rate, Eq.~(\ref{7}). Our approach provides a framework for
systematically studying the mesoscopic phenomena in strongly correlated
electron systems.

We thank D. Maslov, who participated in this work at its early stage,
A.~Altland, V.~Cheianov, T.~Giamarchi, L.~Glazman, P.~Le~Doussal, K.~Le~Hur,
T.~Nattermann, and A.~Tsvelik for interesting discussions. We are particularly
grateful to I.~Aleiner, B.~Altshuler, and D.~Basko for criticism and valuable
comments on the strongly localized regime. The work was supported by SPP
``Quanten-Hall-Systeme" of DFG and by RFBR.


\vspace{-0.5cm}

\begin{thebibliography}{1}

\vspace{-0.6cm}

\bibitem[*]{byline} Also at A.F.~Ioffe Physico-Technical
Institute, 194021 St.~Petersburg, Russia.
 
\bibitem[$\dagger$]{} Also at Petersburg Nuclear Physics
Institute, 188350 St.~Petersburg, Russia.


\bibitem{auslaender02} O.M. Auslaender {\it et al.}, Science {\bf 295}, 825
(2002).

\bibitem{man04} H.T. Man and A.F. Morpurgo, cond-mat/0411141, and references
therein. For transport measurements at room temperature on mm-long diffusive
single-wall nanotubes, see S.~Li, Z.~Yu, C.~Rutherglen, and P.J.~Burke, Nano
Letters, {\bf 4}, 2003 (2004).

\bibitem{kang00} W. Kang {\it et al.}, Nature {\bf 403}, 59 (2000); I.~Yang
{\it et al.}, Phys.\ Rev.\ Lett.\ {\bf 92}, 056802 (2004). 

\bibitem{grayson05} M. Grayson {\it et al.}, Appl.\ Phys.\ Lett.\ {\bf 86},
032101 (2005) and to be published.

\bibitem{altshuler85} B.L. Altshuler and A.G.~Aronov, in {\it
Electron-Electron Interactions in Disordered Systems}, edited by
A.L.~Efros and M.~Pollak (North-Holland, Amsterdam, 1985).
 
\bibitem{giamarchi04} T. Giamarchi, {\it Quantum Physics in One Dimension}
(Oxford University, Oxford, 2004).
                                       
\bibitem{giamarchi88} T. Giamarchi and H.J.~Schulz, Phys.\ Rev.\ B
{\bf 37}, 325 (1988).

\bibitem{furusaki93} A. Furusaki and N. Nagaosa, Phys.\ Rev.\ B {\bf
47}, 4631 (1993); C.L. Kane and M.P.A.~Fisher, Phys.\ Rev.\ Lett.\ {\bf
76}, 3192 (1996); M.-R. Li and E.~Orignac, Europhys.\ Lett.\ {\bf
60}, 432 (2002); T. Giamarchi, T. Nattermann, and P. Le Doussal in {\it
Fundamental Problems of Mesoscopic Physics}, edited by I.V.~Lerner,
B.L.~Altshuler, and Y.~Gefen (Kluwer, Dordrecht, 2004).

\bibitem{apel82a} W. Apel and T.M.~Rice, J. Phys.\ C {\bf 16}, L271
(1982). 
 
\bibitem{lehur02} K. Le Hur, Phys.\ Rev.\ B {\bf 65}, 233314 (2002);
cond-mat/0503652.
 
\bibitem{mattis74} D.C. Mattis, Phys.\ Rev.\ Lett.\ {\bf 32}, 714 (1974);
A. Luther and I. Peschel, {\it ibid.} {\bf 32}, 992 (1974).

\bibitem{polyakov03} See, e.g., D.G. Polyakov and I.V.~Gornyi, Phys.\ Rev.\ B
{\bf 68}, 035421 (2003) and references therein.

\bibitem{unpub} I.V. Gornyi, A.D. Mirlin, and D.G.~Polyakov, to be published.

\bibitem{ludwig04} For diffusive wires, the difference between the WL and
AB dephasing times was shown in T. Ludwig and A.D.~Mirlin, Phys.\
Rev.\ B {\bf 69}, 193306 (2004).

\bibitem{altshuler82} B.L. Altshuler, A.G. Aronov, and D.E.~Khmelnitskii,
J. Phys.\ C {\bf 15}, 7367 (1982).

\bibitem{berezinskii73} V.L. Berezinskii, Sov.\ Phys.\ JETP {\bf 38}, 620
(1974).

\bibitem{geology} In terms of the conventional $g$-ology \cite{giamarchi04}
this means taking $g_4=0$ and $\alpha=g_2/2\pi v_F$.

\bibitem{curvature} Strictly speaking, this is true for a linear dispersion
only. In reality, a finite curvature $m^{-1}$ will lead to an additional
contribution to the dephasing rate $\sim \alpha^2T^2/mv_F^2$. One sees,
however, that Eq.~(\ref{15}) gives the leading contribution in a wide
temperature range $T\ll (m^2v_F^4/\alpha^2\tau)^{1/3}$.

\bibitem{yurkevich} See, e.g., A. Grishin, I.V.~Yurkevich, and I.V.~Lerner,
Phys.\ Rev.\ B {\bf 69}, 165108 (2004) and references therein.


\bibitem{gogolin75} This regime bears resemblance to the phonon-mediated
hopping in A.A. Gogolin, V.I. Mel'nikov, and {\'E}.I. Rashba, Sov.\ Phys.\
JETP {\bf 42}, 168 (1975). Recently, a similar transport mechanism in the
context of a dynamical localization was found in D.M. Basko, Phys.\ Rev.\
Lett.\ {\bf 91}, 206801 (2003).

\bibitem{aleiner05} I.L.~Aleiner, B.L.~Altshuler, and D.M.~Basko, unpublished.

\bibitem{altshuler97} Similar ideas were developed earlier for quantum dots in
B.L. Altshuler, Y. Gefen, A.~Kamenev, and L.S.~Levitov, Phys.\ Rev.\ Lett.\
{\bf 78}, 2803 (1997).  \end{thebibliography}
\end{document}